\documentclass[11pt,a4paper]{article}
\usepackage[T1]{fontenc}
\usepackage[utf8]{inputenc}
\usepackage{amsmath,amssymb,bm,mathtools,booktabs}
\usepackage[expansion=false]{microtype}
\usepackage[margin=25mm]{geometry}
\usepackage{xcolor,graphicx,tikz}
\usetikzlibrary{arrows.meta,positioning,calc,fit}
\usepackage[colorlinks=true,linkcolor=blue!50!black,citecolor=blue!50!black,
 pdftitle={Non-Inertial Response of Correlations: From Scalar Bell Observables to an Extended Correlation Tensor},
 pdfauthor={Timur F. Kamalov},
 pdfsubject={Bell correlations in non-inertial reference frames},
 pdfkeywords={Bell correlations, non-inertial reference frames, correlation tensor, entangled photons, fermionic singlet, quantum tomography}]{hyperref}
\newcommand{\vct}[1]{\bm{#1}}
\newcommand{\Corr}{\mathcal{C}}
\newcommand{\FullT}{\mathcal{T}}
\newcommand{\Tr}{\operatorname{Tr}}
\newcommand{\res}{\mathrm{res}}

\newcommand{\ket}[1]{\lvert #1\rangle}
\title{Non-Inertial Response of Correlations:\\
From Scalar Bell Observables to an Extended Correlation Tensor}
\author{Timur F. Kamalov\\
\small State University of Education, Moscow, Russia\\
\small \texttt{timkamalov@gmail.com}}
\date{}

\begin{document}
\maketitle
\begin{center}
\small Preprint available at Zenodo:
\href{https://doi.org/10.5281/zenodo.21932616}
{https://doi.org/10.5281/zenodo.21932616}
\end{center}

\begin{abstract}
A standard Bell observable is a scalar correlation associated with a selected
pair of local measurement directions. We formulate it as a projection of a
complete two-particle correlation tensor and distinguish two angular sectors.
The central result is a reversal of the sign multiplying the cosine law: for
coincident calibrated settings, the Bell observable is positive for photons
and negative for a fermionic singlet. This distinction provides an operational
criterion for experimental identification. For photons, the positive sign
follows from averaging two projection amplitudes over the complete
non-inertial phase interval; the fermionic sign follows from the negative
exchange holonomy of the phase--momentum sector. Mapping phase directions to
linear-polarizer axes produces the corresponding double-angle dependence. The
photon Stokes tensor has positive linear-polarization components and a negative
circular-polarization component, whereas the fermionic singlet has an
isotropic negative tensor. We introduce a motion-dependent extended tensor and
a frequency-dependent non-inertial susceptibility. A phase-synchronous
two-arm experiment combines optical modulation, rotation, and seeded
multiaxial piezoelectric vibration. Independent motion measurements separate
common and differential components and allow controlled variation from
correlated to independent and oppositely driven motion; a single rigid
platform is the simpler common-frame limit. Bell-setting and state-correlation
vectors express the sign reversal as a scalar projection. The construction
yields binary joint probabilities and recovers the Tsirelson bound with
oppositely signed optimal CHSH combinations.

\end{abstract}


\section{Introduction}

In a standard Bell experiment~\cite{Bell1964,CHSH1969}, the correlation between two binary outcomes is
represented by a number $E(\vct a,\vct b)$ for selected analyzer directions
$\vct a$ and $\vct b$. Several such numbers form the CHSH combination. A
single Bell observable, however, is not a complete description of the
two-particle correlation structure; it is only one scalar projection of that
structure.

This distinction becomes important when the local measurement systems rotate,
accelerate, or undergo controlled non-inertial modulation. A variation of the
recorded correlation may result from a change of the local measurement frames,
from propagation and mode transformations, or from a response of the state and
its intrinsic correlation structure. A single value of the CHSH parameter
does not separate these contributions.

Our aim is to formulate Bell observables as projections of an extended
correlation tensor and to define an experimentally reconstructible response of
that tensor to non-inertial modulation. The ordinary two-qubit correlation
matrix is well known in quantum-information theory~\cite{Horodecki1995}. The additional object introduced
here is its non-inertial susceptibility, including the amplitude and phase of
the response to a periodic perturbation.

\section{Photon Correlation on the Complete Phase Interval}

In the non-inertial description, the complete state of a pair may include a
common phase variable $\lambda$ and a random metric structure. Let
$A(a,\lambda)$ and $B(b,\lambda)$ denote local projection amplitudes. The
physical axis of linear photon polarization is projective and has period
$\pi$, but the phase variable of the correlation state completes the full
cycle $0\leq\lambda<2\pi$. We therefore define
\begin{equation}
 M_{AB}^{(\gamma)}(a,b)=\frac{1}{\pi}\int_0^{2\pi}
 A(a,\lambda)B(b,\lambda)\,d\lambda.
 \label{eq:photon_full_phase_average}
\end{equation}
For the projection amplitudes
\begin{equation}
 A(a,\lambda)=\cos(a-\lambda),\qquad
 B(b,\lambda)=\cos(b-\lambda),
 \label{eq:photon_projection_amplitudes}
\end{equation}
the identity $\cos x\cos y=\tfrac12[\cos(x-y)+\cos(x+y)]$ gives
\begin{equation}
 M_{AB}^{(\gamma)}(a,b)=+\cos(a-b).
 \label{eq:positive_photon_correlation}
\end{equation}
Thus coincident phase directions are positively correlated,
$M_{AB}^{(\gamma)}(a,a)=+1$. The integration interval belongs to the phase
of the complete correlation state and must not be confused with the
projective periodicity of a linear-polarizer axis. On the equator of the
Poincare sphere, $a=2\theta_A$ and $b=2\theta_B$, and hence
\begin{equation}
 E_\gamma(\theta_A,\theta_B)=+\cos2(\theta_A-\theta_B).
 \label{eq:positive_photon_polarizer_correlation}
\end{equation}
This positive photon sector is the baseline used below. It is distinct from
the negative correlation of a fermionic spin singlet.

\section{Exchange Holonomy and the Fermionic Sign}

The complete non-inertial state is defined in an extended phase space
$\Gamma_{\rm ni}=(\lambda,p_\lambda)$. For two indistinguishable particles,
an exchange is represented by a closed path $C_{\rm ex}$ in the reduced
two-particle configuration space. Such a space is multiply connected and its
path integral admits the two one-dimensional exchange factors $+1$ and $-1$
\cite{LaidlawDeWitt1971}. We write the corresponding holonomy as
\begin{equation}
 \mathcal H_\varepsilon(C_{\rm ex})=
 \exp\!\left(i\oint_{C_{\rm ex}}\mathcal A\right)=\varepsilon,
 \qquad \varepsilon=\pm1,
 \label{eq:exchange_holonomy}
\end{equation}
where $\mathcal A$ is the effective connection of the transported extended
state. The symmetric photon sector carries $\varepsilon_\gamma=+1$, whereas
the antisymmetric fermionic sector carries $\varepsilon_f=-1$. This statement
does not attempt to rederive the relativistic spin--statistics theorem; it
identifies the exchange character retained by the complete non-inertial
phase--momentum sector.

Transport of the second local projection around the exchange loop gives
\begin{equation}
 A_\varepsilon(a,\lambda)=\cos(a-\lambda),\qquad
 B_\varepsilon(b,\lambda)=\varepsilon\cos(b-\lambda).
 \label{eq:exchange_projectors}
\end{equation}
Consequently,
\begin{equation}
 M_{AB}^{(\varepsilon)}(a,b)=\frac1\pi\int_0^{2\pi}
 A_\varepsilon(a,\lambda)B_\varepsilon(b,\lambda)\,d\lambda
 =\varepsilon\cos(a-b).
 \label{eq:unified_signed_correlation}
\end{equation}
Thus the positive photon and negative fermionic cosine laws follow from one
phase average and the two exchange holonomies:
\begin{equation}
 M_{AB}^{(\gamma)}=+\cos(a-b),\qquad
 M_{AB}^{(f)}=-\cos(a-b).
 \label{eq:photon_fermion_sign_reversal}
\end{equation}
The standard Bell states used later are representations of these two tensor
sectors, rather than the origin of the signs in Eq.~\eqref{eq:photon_fermion_sign_reversal}.

\section{Bell-Setting and Correlation Vectors}

The signed phase average above has a simple two-dimensional representation.
Let
\begin{equation}
 \Delta=a-b
 \label{eq:relative_phase_delta}
\end{equation}
be the relative phase direction of the two analyzers. Introduce two
quadratures in arm $A$ and one reference quadrature in arm $B$,
\begin{equation}
 A_c(a,\lambda)=\cos(a-\lambda),\qquad
 A_s(a,\lambda)=\sin(a-\lambda),\qquad
 B_c(b,\lambda)=\cos(b-\lambda).
 \label{eq:phase_quadratures}
\end{equation}
Their normalized overlaps over the complete photon phase interval are
\begin{align}
 \frac1\pi\int_0^{2\pi}A_c(a,\lambda)B_c(b,\lambda)\,d\lambda
 &=\cos\Delta,\notag\\
 \frac1\pi\int_0^{2\pi}A_s(a,\lambda)B_c(b,\lambda)\,d\lambda
 &=\sin\Delta.
 \label{eq:quadrature_overlaps}
\end{align}
The factor $1/\pi$ is the normalization weight of the correlation overlap;
it is not a separately normalized probability density for $\lambda$.
Equation~\eqref{eq:quadrature_overlaps} motivates the Bell-setting vector
\begin{equation}
 \vct b_{AB}(\Delta)=
 \begin{pmatrix}\cos\Delta\\ \sin\Delta\end{pmatrix}
 =\frac1\pi\int_0^{2\pi}
 \begin{pmatrix}A_c(a,\lambda)B_c(b,\lambda)\\
 A_s(a,\lambda)B_c(b,\lambda)
 \end{pmatrix}d\lambda.
 \label{eq:bell_setting_vector}
\end{equation}
This vector is not a spatial polarization or spin direction. It is an
auxiliary vector of two angular functions fixed only by the relative analyzer
setting.

Equivalently, with the local phase vectors
\begin{equation}
 \vct n_A(a)=\begin{pmatrix}\cos a\\\sin a\end{pmatrix},
 \qquad
 \vct n_B(b)=\begin{pmatrix}\cos b\\\sin b\end{pmatrix},
\end{equation}
one may write
\begin{equation}
 \vct b_{AB}=
 \begin{pmatrix}
 \vct n_A\!\cdot\!\vct n_B\\
 \vct e_z\!\cdot\!(\vct n_B\!\times\!\vct n_A)
 \end{pmatrix}
 =\begin{pmatrix}\cos(a-b)\\\sin(a-b)\end{pmatrix}.
 \label{eq:dot_cross_bell_vector}
\end{equation}
Thus its cosine component is a scalar product and its sine component is the
oriented cross product in the analyzer plane.

The state of the pair is described independently by the correlation vector
\begin{equation}
 \vct c=V\begin{pmatrix}\cos\varphi\\\sin\varphi\end{pmatrix},
 \qquad 0\leq V\leq1,
 \label{eq:state_correlation_vector}
\end{equation}
where $V=|\vct c|$ is the correlation visibility and $\varphi$ is the phase
orientation of the correlation pattern. In the ideal, fully correlated case
$V=1$; values $0\leq V<1$ describe reduced visibility caused by imperfect
state preparation, decoherence, background counts, or other experimental
losses of correlation contrast. The scalar Bell observable is then
\begin{equation}
 E(\Delta)=\vct c\cdot\vct b_{AB}(\Delta)
 =V\cos(\Delta-\varphi).
 \label{eq:vector_bell_observable}
\end{equation}
This definition is not circular: $(V,\varphi)$ specify the prepared state,
$\Delta$ specifies the apparatus, and Eq.~\eqref{eq:vector_bell_observable}
predicts the measured scalar. Estimating $\vct c$ from measured values of $E$
is the inverse experimental problem.

For the two ideal exchange sectors, Eq.~\eqref{eq:unified_signed_correlation}
corresponds to
\begin{equation}
 \vct c_\varepsilon=
 \begin{pmatrix}\varepsilon\\0\end{pmatrix},
 \qquad
 E_\varepsilon(\Delta)=\vct c_\varepsilon\cdot\vct b_{AB}(\Delta)
 =\varepsilon\cos\Delta.
 \label{eq:signed_vector_proof}
\end{equation}
For linear photon polarization,
\begin{equation}
 \Delta_\gamma=2(\theta_A-\theta_B),\qquad
 \vct c_\gamma=\begin{pmatrix}+1\\0\end{pmatrix},
\end{equation}
and therefore
\begin{equation}
 E_\gamma(\theta_A,\theta_B)
 =\vct c_\gamma\cdot\vct b_{AB}(\Delta_\gamma)
 =+\cos2(\theta_A-\theta_B),
 \qquad E_\gamma(0)=+1>0.
 \label{eq:photon_vector_proof}
\end{equation}
For coplanar fermionic spin analyzers,
\begin{equation}
 \Delta_f=\alpha_A-\alpha_B,\qquad
 \vct c_f=\begin{pmatrix}-1\\0\end{pmatrix},
\end{equation}
and hence
\begin{equation}
 E_f(\alpha_A,\alpha_B)
 =\vct c_f\cdot\vct b_{AB}(\Delta_f)
 =-\cos(\alpha_A-\alpha_B),
 \qquad E_f(0)=-1<0.
 \label{eq:fermion_vector_proof}
\end{equation}
Equations~\eqref{eq:photon_vector_proof} and
\eqref{eq:fermion_vector_proof} are two vector realizations of the same Bell
observable, not two different definitions of it. The sign is meaningful only
after the output-channel labels and the coincident settings have been
calibrated consistently.

\section{Binary Bell Probabilities and the Tsirelson Bound}

Let the local outcomes be $A,B=\pm1$. For fixed settings,
\begin{equation}
 E(\vct a,\vct b)=\sum_{A,B=\pm1}AB\,P(A,B\mid\vct a,\vct b).
 \label{eq:bell_probability}
\end{equation}
For two-level systems, the local observables are
\begin{equation}
 \hat A(\vct a)=a_i\hat\sigma_i,
 \qquad \hat B(\vct b)=b_j\hat\sigma_j,
\end{equation}
and hence
\begin{equation}
 E(\vct a,\vct b)=\Tr\!\left[\hat\rho\,
 \hat A(\vct a)\otimes\hat B(\vct b)\right].
\end{equation}
The CHSH parameter is
\begin{equation}
 S=E(\vct a,\vct b)+E(\vct a,\vct b')
 +E(\vct a',\vct b)-E(\vct a',\vct b').
 \label{eq:chsh}
\end{equation}
It combines four scalar projections but does not reconstruct the complete
tensor structure.

For unbiased local marginals, the signed non-inertial correlation
$E_\varepsilon(a,b)=\varepsilon\cos(a-b)$ determines the binary joint
probabilities
\begin{align}
 P_{++}=P_{--}&=\frac14[1+\varepsilon\cos(a-b)],\notag\\
 P_{+-}=P_{-+}&=\frac14[1-\varepsilon\cos(a-b)].
 \label{eq:binary_joint_probabilities}
\end{align}
They are nonnegative, normalized, and reproduce the Bell observable,
\begin{equation}
 \sum_{r,s=\pm1}rs\,P_{rs}(a,b)=\varepsilon\cos(a-b).
 \label{eq:binary_bridge}
\end{equation}
Equation~\eqref{eq:binary_bridge} supplies the explicit bridge from the
continuous projection amplitudes to binary coincidence outcomes.

The same construction directly recovers the Tsirelson bound
\cite{Cirelson1980}. Introduce the unit phase vector
$\vct n(a)=(\cos a,\sin a)$, so that
$E_\varepsilon(a,b)=\varepsilon\vct n(a)\cdot\vct n(b)$. The CHSH
combination becomes
\begin{align}
 S_\varepsilon=\varepsilon\{&\vct n(a)\cdot[\vct n(b)+\vct n(b')]
 +\vct n(a')\cdot[\vct n(b)-\vct n(b')]\}.
 \label{eq:chsh_phase_vectors}
\end{align}
Maximization over $a$ and $a'$ gives
\begin{equation}
 |S_\varepsilon|\leq
 |\vct n(b)+\vct n(b')|+|\vct n(b)-\vct n(b')|.
\end{equation}
If the angle between $\vct n(b)$ and $\vct n(b')$ is $\delta$, then
\begin{equation}
 |S_\varepsilon|\leq2\left(
 \left|\cos\frac\delta2\right|+
 \left|\sin\frac\delta2\right|\right)\leq2\sqrt2.
 \label{eq:tsirelson_from_phase}
\end{equation}
For $a=0$, $a'=\pi/2$, $b=\pi/4$, and $b'=-\pi/4$, the two sectors give
\begin{equation}
 S_\gamma=+2\sqrt2,\quad S_f=-2\sqrt2,\quad
 |S_\gamma|=|S_f|=2\sqrt2.
 \label{eq:opposite_signed_tsirelson}
\end{equation}
Thus the exchange sign reverses the oriented CHSH combination without
changing its maximal absolute magnitude. For physical photon polarizers the
phase angles are doubled, giving the familiar settings $0^\circ$, $45^\circ$,
$22.5^\circ$, and $-22.5^\circ$.

\section{Bell Observable as a Scalar Projection}

Define the two-qubit correlation matrix
\begin{equation}
 C_{ij}=\Tr\!\left[\hat\rho
 (\hat\sigma_i\otimes\hat\sigma_j)\right]
 =\left\langle\hat\sigma_i\otimes\hat\sigma_j\right\rangle.
 \label{eq:correlation_tensor}
\end{equation}
Then
\begin{equation}
 E(\vct a,\vct b)=a_iC_{ij}b_j=\vct a^{\mathsf T}C\vct b.
 \label{eq:scalar_projection}
\end{equation}
The variation of the measured scalar separates three contributions:
\begin{equation}
 \delta E=(\delta\vct a)^{\mathsf T}C\vct b
 +\vct a^{\mathsf T}(\delta C)\vct b
 +\vct a^{\mathsf T}C(\delta\vct b).
 \label{eq:variation_E}
\end{equation}
The first and third terms describe changes of the local directions; the middle
term describes a change of the correlation matrix itself. In orthonormal local
bases, its elements are obtained from
\begin{equation}
 C_{ij}=E(\vct e_i,\vct e_j).
\end{equation}

\subsection{Complete extended tensor}

Set $\hat\sigma_0=I$ and define
\begin{equation}
 \FullT_{\mu\nu}=\Tr\!\left[\hat\rho
 (\hat\sigma_\mu\otimes\hat\sigma_\nu)\right],
 \qquad \mu,\nu=0,1,2,3.
 \label{eq:full_extended_tensor}
\end{equation}
It has the block form
\begin{equation}
 \FullT=\begin{pmatrix}1&\vct v^{\mathsf T}\\ \vct u&C\end{pmatrix},
 \qquad
 u_i=\langle\hat\sigma_i\otimes I\rangle,
 \quad v_j=\langle I\otimes\hat\sigma_j\rangle.
 \label{eq:full_tensor_blocks}
\end{equation}
The vectors $\vct u$ and $\vct v$ are the local polarizations, while $C$ is the
joint correlation block. The complete joint probability is
\begin{equation}
 P(A,B\mid\vct a,\vct b)=\frac14\left[
 1+A\vct a\cdot\vct u+B\vct b\cdot\vct v
 +ABa_iC_{ij}b_j\right].
 \label{eq:joint_probability}
\end{equation}
Thus, a Bell observable probes only the correlation block, whereas the single
counts also determine the local blocks of the full tensor.

\section{Extended Non-Inertial Correlation Tensor}

Let $\xi^\alpha$ denote controlled motion parameters, including orientation,
angular velocity, acceleration, jerk, modulation phase, or stochastic
variables. We introduce
\begin{equation}
 \Corr_{ij}(\vct\xi)=\Tr\!\left[
 \hat\rho(\vct\xi)\,
 \hat\Sigma_i^{(A)}(\vct\xi)\otimes
 \hat\Sigma_j^{(B)}(\vct\xi)\right],
 \label{eq:extended_tensor}
\end{equation}
and the corresponding scalar projection
\begin{equation}
 E_{\rm ni}(\vct a,\vct b;\vct\xi)
 =a_i\Corr_{ij}(\vct\xi)b_j.
\end{equation}
The complete tensor is
\begin{equation}
 \FullT(\vct\xi)=
 \begin{pmatrix}1&\vct v^{\mathsf T}(\vct\xi)\\
 \vct u(\vct\xi)&\Corr(\vct\xi)\end{pmatrix}.
 \label{eq:full_noninertial_tensor}
\end{equation}

To separate motion of the measurement bases from an intrinsic response, write
\begin{equation}
 \Corr_{\rm lab}(\vct\xi)=R_A(\vct\xi)
 \Corr_{\rm int}(\vct\xi)R_B^{\mathsf T}(\vct\xi).
 \label{eq:lab_intrinsic}
\end{equation}
After compensation of the calibrated frame rotations, the residual tensor is
\begin{equation}
 \Delta\Corr_{\res}(\vct\xi)=R_A^{-1}\Corr_{\rm lab}
 R_B^{-\mathsf T}-C^{(0)}.
 \label{eq:residual_tensor}
\end{equation}
A nonzero residual is not by itself evidence for new physics: decoherence,
channel imperfections, drift, and standard relativistic transformations must
first be included in the baseline prediction.

\section{Photon Polarization}

For photons, use normalized Stokes operators,
\begin{equation}
 C_{\mu\nu}^{(\gamma)}=
 \langle\hat S_\mu^{(A)}\otimes\hat S_\nu^{(B)}\rangle,
 \qquad \mu,\nu=1,2,3.
\end{equation}
With $\hat S_0=I$, the complete two-photon Stokes tensor has indices
$\mu,\nu=0,1,2,3$. The positive linear-polarization correlation is realized
by
\begin{equation}
 \ket{\Phi^+}=\frac1{\sqrt2}(\ket H_A\ket H_B+\ket V_A\ket V_B).
 \label{eq:photon_phi_plus}
\end{equation}
In the Stokes-axis order $(H/V,D/A,R/L)$,
\begin{equation}
 C^{(\gamma,0)}=\operatorname{diag}(1,1,-1),\qquad
 \FullT^{(\gamma,0)}=\operatorname{diag}(1,1,1,-1).
 \label{eq:photon_positive_tensor}
\end{equation}
The vanishing local marginals express the absence of polarization of either
photon separately. The two positive entries describe positive correlations of
linear polarization; the circular component has the sign fixed by the chosen
Bell state and Stokes-axis convention. A linear analyzer at angle $\theta$
corresponds to
\begin{equation}
 \vct n(\theta)=(\cos2\theta,\sin2\theta,0),
\end{equation}
and therefore
\begin{equation}
 E_0^{(\gamma)}(\theta_A,\theta_B)=
 \vct n^{\mathsf T}(\theta_A)C^{(\gamma,0)}\vct n(\theta_B)
 =+\cos2(\theta_A-\theta_B).
 \label{eq:photon_angular_correlation}
\end{equation}
The double angle expresses the equivalence of the physical axes $\theta$ and
$\theta+\pi$; it does not reduce the full integration interval of the phase
variable $\lambda$. The complete $3\times3$ block requires linear, diagonal,
and circular polarization projections. A fit to a single cosine law does not
replace tensor reconstruction.

\section{Fermionic Singlet}

For
\begin{equation}
 \ket{\Psi^-}=\frac1{\sqrt2}
 (\ket{\uparrow\downarrow}-\ket{\downarrow\uparrow}),
\end{equation}
one has
\begin{equation}
 C_{ij}^{(f,0)}=-\delta_{ij},\qquad
 \FullT^{(f,0)}=\operatorname{diag}(1,-1,-1,-1).
 \label{eq:fermion_singlet_tensor}
\end{equation}
For coplanar spin analyzers,
\begin{equation}
 E_0^{(f)}(\alpha_A,\alpha_B)
 =-\vct a\cdot\vct b=-\cos(\alpha_A-\alpha_B).
 \label{eq:fermion_angular_correlation}
\end{equation}
The single angle for spin, compared with the double angle for linear photon
polarization, reflects the difference between an oriented spin direction and
an unoriented polarization axis.

In a non-inertial setting,
\begin{equation}
 \FullT^{(f)}(\vct\xi)=
 \begin{pmatrix}1&\vct v^{\mathsf T}(\vct\xi)\\
 \vct u(\vct\xi)&-I+\Delta C^{(f)}(\vct\xi)\end{pmatrix}.
\end{equation}
For relativistic massive particles, the spin transformation is momentum
dependent. At fixed momenta,
\begin{equation}
 C_{\rm lab}^{(f)}=R_W^{(A)}(\vct p_A)
 C_{\rm int}^{(f)}R_W^{(B)\mathsf T}(\vct p_B),
 \label{eq:wigner_tensor}
\end{equation}
where $R_W$ denotes the Wigner rotation. If momenta are unresolved, the
observed tensor is averaged over the joint distribution,
\begin{equation}
 \overline C_{ij}^{(f)}=\int d^3p_A\,d^3p_B\,
 f(\vct p_A,\vct p_B)C_{ij}^{(f)}(\vct p_A,\vct p_B).
 \label{eq:momentum_average}
\end{equation}
Wigner rotations and spin--momentum averaging belong to the standard
relativistic background and must be compensated before a residual
non-inertial response is inferred~\cite{GingrichAdami2002,PeresTerno2004,TerashimaUeda2003}.

\section{Non-Inertial Susceptibility}

Near the unperturbed state,
\begin{equation}
 \Corr_{ij}(\vct\xi)=C_{ij}^{(0)}+\chi_{ij\alpha}\xi^\alpha
 +\frac12\chi^{(2)}_{ij\alpha\beta}\xi^\alpha\xi^\beta+\cdots,
\end{equation}
with
\begin{equation}
 \chi_{ij\alpha}=\left.
 \frac{\partial\Corr_{ij}}{\partial\xi^\alpha}\right|_{\vct\xi=0}.
 \label{eq:susceptibility}
\end{equation}
For the full tensor,
\begin{equation}
 X_{\mu\nu\alpha}=\left.
 \frac{\partial\FullT_{\mu\nu}}{\partial\xi^\alpha}
 \right|_{\vct\xi=0},
\end{equation}
where $X_{i0\alpha}$ and $X_{0j\alpha}$ describe induced local polarization
and $X_{ij\alpha}=\chi_{ij\alpha}$ describes the correlation response.

For harmonic modulation, $\xi^\alpha(t)=\xi_0^\alpha\cos\Omega t$, a delayed
response is described by a complex susceptibility,
\begin{equation}
 \delta\Corr_{ij}(\Omega)=
 \chi_{ij\alpha}(\Omega)\xi^\alpha(\Omega).
 \label{eq:dynamic_susceptibility}
\end{equation}
Its real and imaginary parts represent in-phase and quadrature responses. The
observable scalar susceptibility is
\begin{equation}
 \chi_E^{(\alpha)}(\vct a,\vct b;\Omega)
 =a_i\chi_{ij\alpha}(\Omega)b_j.
\end{equation}

\section{Phenomenological Response Structure}

The fermionic and photon baselines must be expanded separately. For the
isotropic fermionic singlet, a minimal residual correction is
\begin{equation}
 \Delta C_{ij}^{(f),\res}=-\eta_f\delta_{ij}
 +q_f\left(m_im_j-l_il_j\right)
 +\kappa_f\epsilon_{ijk}n_k,
 \label{eq:fermion_phenomenological_decomposition}
\end{equation}
where $\eta_f$ changes the overall spin anticorrelation, $q_f$ describes a
symmetric anisotropy, and $\kappa_f$ is an antisymmetric rotational response.

For photons propagating parallel to the rotation axis,
$\vct n\parallel\vct k$, choose transverse Stokes basis vectors
$\vct e_1=\vct m$ and $\vct e_2=\vct l$. The unperturbed linear block is
$+I_2$, not $-I_2$. Its minimal response is therefore
\begin{equation}
 \Corr_\perp^{(\gamma)}(\Omega)=
 \begin{pmatrix}
 1-\eta_\gamma+q_\gamma&-\kappa_\gamma\\
 \kappa_\gamma&1-\eta_\gamma-q_\gamma
 \end{pmatrix}.
 \label{eq:transverse_matrix}
\end{equation}
Here $\eta_\gamma$ describes a reduction of the positive photon correlation,
$q_\gamma$ is the symmetric linear-polarization anisotropy, and
$\kappa_\gamma$ is the antisymmetric rotational component. A nonzero
$\kappa_\gamma$ can imitate a small relative analyzer rotation; changes of
$\eta_\gamma$, $q_\gamma$, or of the singular values after kinematic
compensation provide stronger evidence for a modified correlation structure.
At low frequency one may write
\begin{align}
 \eta_\gamma(\Omega)&=\eta_{\gamma0}+\eta_{\gamma2}\Omega^2+\cdots,\notag\\
 q_\gamma(\Omega)&=q_{\gamma0}+q_{\gamma2}\Omega^2+\cdots,\notag\\
 \kappa_\gamma(\Omega)&=i\kappa_{\gamma1}\Omega+i\kappa_{\gamma3}\Omega^3+\cdots.
 \label{eq:low_frequency_expansion}
\end{align}
This is a phenomenological tensor and frequency structure, not a numerical
prediction. A microscopic theory must determine which coefficients are
nonzero and their magnitude.

\section{Proposed Photon Experiment}

\subsection{Operational summary}

The primary data are single counts and two-photon coincidences, time-tagged
relative to the mechanical modulation phase. For each analyzer pair, record
$N_{++},N_{+-},N_{-+},N_{--}$ and compute
\begin{equation}
 C=\frac{N_{++}+N_{--}-N_{+-}-N_{-+}}
 {N_{++}+N_{--}+N_{+-}+N_{-+}}.
 \label{eq:count_correlation}
\end{equation}
Single counts determine the local polarizations,
\begin{equation}
 u_i=\frac{N_{+,i}^{(A)}-N_{-,i}^{(A)}}
 {N_{+,i}^{(A)}+N_{-,i}^{(A)}},\qquad
 v_j=\frac{N_{+,j}^{(B)}-N_{-,j}^{(B)}}
 {N_{+,j}^{(B)}+N_{-,j}^{(B)}}.
\end{equation}
Complete tomography uses the $H/V$, $D/A$, and $R/L$ bases. A first
proof-of-principle test may use only
$C_{11},C_{22},C_{12},C_{21}$, from which
\begin{equation}
 \eta_\gamma=1-\frac{C_{11}+C_{22}}2,\qquad
 q_\gamma=\frac{C_{11}-C_{22}}2,\qquad
 \kappa_\gamma=\frac{C_{21}-C_{12}}2.
 \label{eq:extract_coefficients}
\end{equation}

\subsection{Optical arrangement and modulation}

An SPDC source prepares the positively correlated Bell state
\begin{equation}
 \ket{\Phi^+}=\frac1{\sqrt2}
 (\ket H_A\ket H_B+\ket V_A\ket V_B),
\end{equation}
for which $E_0^{(\gamma)}=+\cos2(\theta_A-\theta_B)$. In phase variables
$a=2\theta_A$ and $b=2\theta_B$, this is precisely
$M_{AB}^{(\gamma)}=+\cos(a-b)$ obtained from the complete phase interval.

The two photons are analyzed in modules $A$ and $B$ that share the same
controlled non-inertial motion. Wherever practicable, both modules are mounted
on one rigid moving platform. An equivalent implementation uses two matched
stages driven synchronously and monitored independently. For harmonic
rotation,
\begin{equation}
 \phi_A(t)\simeq\phi_B(t)=\phi_0\cos\Omega t.
\end{equation}
It is useful to separate the common and differential coordinates,
\begin{equation}
 \phi_+(t)=\frac{\phi_A(t)+\phi_B(t)}2,
 \qquad
 \phi_-(t)=\phi_A(t)-\phi_B(t).
 \label{eq:common_differential_rotation}
\end{equation}
The target common-frame experiment has $\phi_-(t)\simeq0$. For the common
coordinate $\phi_+(t)=\phi_0\cos\Omega t$,
\begin{align}
 \dot\phi_+&=-\Omega\phi_0\sin\Omega t,\notag\\
 \ddot\phi_+&=-\Omega^2\phi_0\cos\Omega t,\notag\\
 \dddot\phi_+&=\Omega^3\phi_0\sin\Omega t.
\end{align}
Measurements at several $\Omega$ and $\phi_0$ separate different frequency
scalings. Phenomenologically,
\begin{equation}
 \delta\Corr_{ij}(\Omega)=
 [\chi_{ij}^{(\phi,+)}-i\Omega\chi_{ij}^{(\dot\phi,+)}
 -\Omega^2\chi_{ij}^{(\ddot\phi,+)}
 +i\Omega^3\chi_{ij}^{(\dddot\phi,+)}]\phi_+(\Omega)
 +\chi_{ij}^{(\phi,-)}(\Omega)\phi_-(\Omega).
 \label{eq:frequency_scaling}
\end{equation}
The term proportional to $\phi_-$ measures imperfect synchronization and the
ordinary differential basis response. It must be separated from the common
response. Indeed, the standard photon law is unchanged by an equal rotation
of the two analyzer axes,
\begin{equation}
 \cos2[(\theta_A+\phi_+)-(\theta_B+\phi_+)]
 =\cos2(\theta_A-\theta_B).
 \label{eq:common_rotation_null}
\end{equation}
Thus common rotation supplies a particularly clean null geometry: after
calibration, any residual dependence on $\dot\phi_+$, $\ddot\phi_+$, or higher
non-inertial variables cannot be attributed to a change of the relative
analyzer angle.

Rotation is retained as a calibrated reference because it generates a known
change of orientation together with angular velocity and angular acceleration.
It is complemented by a translational vibration mode in which the orientation
of both analyzers is kept fixed while modules $A$ and $B$ are mounted on a
common multiaxial stage or on two synchronized two- or three-axis
piezoelectric stages. A computer generates the same reproducible seeded,
band-limited pseudorandom drive signals $u_k(t)$ for the two stages. Equal
drive voltages alone do not establish equal motion because the stages,
mountings, and payloads can have different transfer functions. Therefore a
calibrated accelerometer is attached to each module and records
\begin{align}
 a_{A,k}(\omega)&=H_{A,k}(\omega)u_k(\omega),\notag\\
 a_{B,k}(\omega)&=H_{B,k}(\omega)u_k(\omega),
 \label{eq:piezo_transfer}
\end{align}
where $H_{A,k}$ and $H_{B,k}$ are the measured electromechanical transfer
functions. Feedback or digital precompensation is used to make the measured
accelerations coincide. Define
\begin{equation}
 \vct a_+(t)=\frac{\vct a_A(t)+\vct a_B(t)}2,
 \qquad
 \vct a_-(t)=\vct a_A(t)-\vct a_B(t).
 \label{eq:common_differential_acceleration}
\end{equation}
The common-frame condition is $\vct a_-(t)\simeq0$. A piezoelectric actuator
cannot sustain uniform acceleration over an unlimited time because its
displacement is bounded; it instead supplies zero-mean harmonic or stochastic
acceleration over a controlled frequency band.

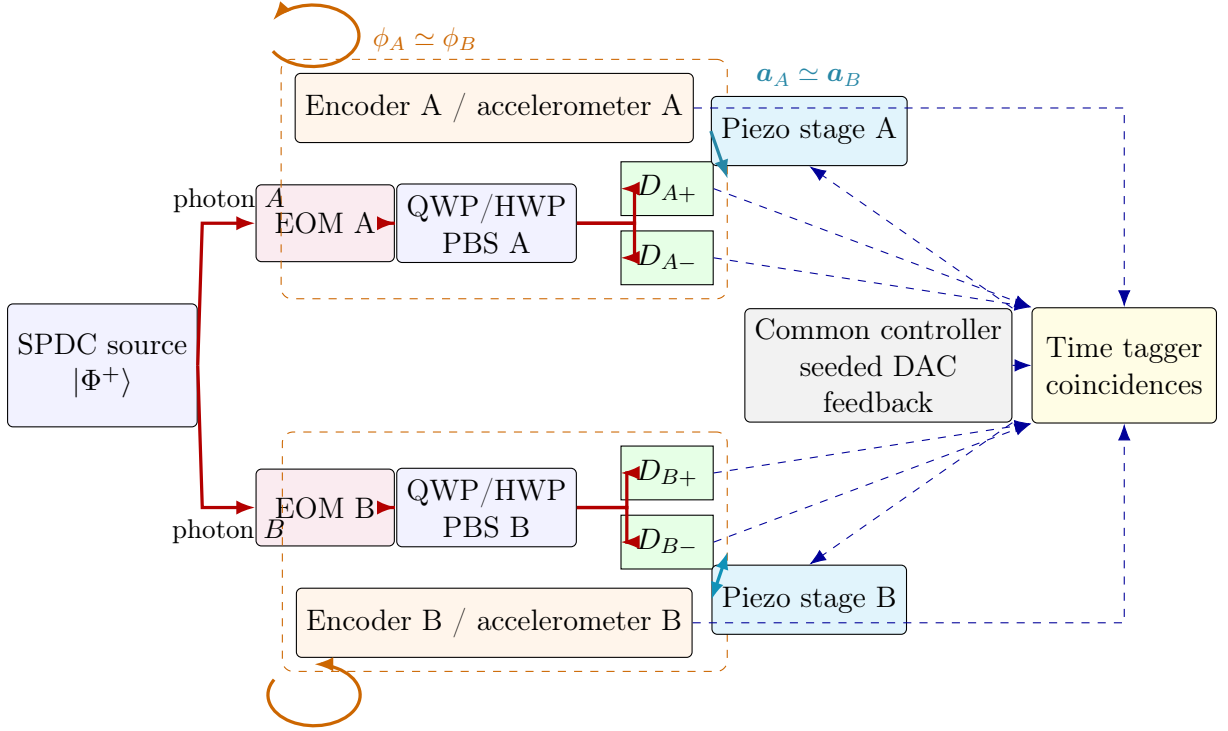
\begin{figure}[ht]
\centering
\resizebox{\textwidth}{!}{%
\begin{tikzpicture}[>=Latex,
beam/.style={draw=red!70!black,very thick,-{Latex[length=2.5mm]}},
signal/.style={draw=blue!60!black,dashed,-{Latex[length=2.2mm]}},
block/.style={draw,rounded corners=2pt,minimum width=18mm,minimum height=10mm,align=center,fill=blue!5},
detector/.style={draw,minimum width=12mm,minimum height=7mm,align=center,fill=green!10},
platform/.style={draw=orange!80!black,dashed,rounded corners=4pt,inner sep=5pt},
note/.style={align=center,font=\small}]
\node[block,minimum width=23mm,minimum height=16mm] (src) at (0,0)
 {SPDC source\\$\ket{\Phi^+}$};
\coordinate (splitA) at (1.3,1.85); \coordinate (splitB) at (1.3,-1.85);
\node[block,fill=purple!8] (eomA) at (2.9,1.85) {EOM A};
\node[block] (anaA) at (5.0,1.85) {QWP/HWP\\PBS A};
\node[detector] (Ap) at (7.35,2.30) {$D_{A+}$};
\node[detector] (Am) at (7.35,1.40) {$D_{A-}$};
\node[block,fill=purple!8] (eomB) at (2.9,-1.85) {EOM B};
\node[block] (anaB) at (5.0,-1.85) {QWP/HWP\\PBS B};
\node[detector] (Bp) at (7.35,-1.40) {$D_{B+}$};
\node[detector] (Bm) at (7.35,-2.30) {$D_{B-}$};
\node[block,minimum width=27mm,minimum height=9mm,fill=orange!8]
 (sensA) at (5.1,3.35) {Encoder A / accelerometer A};
\node[block,minimum width=27mm,minimum height=9mm,fill=orange!8]
 (sensB) at (5.1,-3.35) {Encoder B / accelerometer B};
\node[platform,fit=(anaA)(Ap)(Am)(sensA)] (movingA) {};
\node[platform,fit=(anaB)(Bp)(Bm)(sensB)] (movingB) {};
\node[block,minimum width=23mm,minimum height=9mm,fill=cyan!10]
 (piezoA) at (9.2,3.05) {Piezo stage A};
\node[block,minimum width=23mm,minimum height=9mm,fill=cyan!10]
 (piezoB) at (9.2,-3.05) {Piezo stage B};
\node[block,minimum width=28mm,minimum height=12mm,fill=gray!10]
 (rng) at (10.1,0) {Common controller\\seeded DAC\\feedback};
\node[block,minimum width=24mm,minimum height=15mm,fill=yellow!12]
 (tag) at (13.3,0) {Time tagger\\coincidences};
\draw[beam] (src.east)--(splitA)--node[above,note]{photon $A$}(eomA.west);
\draw[beam] (eomA.east)--(anaA.west);
\draw[beam] (anaA.east)--++(0.75,0)|-(Ap.west);
\draw[beam] (anaA.east)--++(0.75,0)|-(Am.west);
\draw[beam] (src.east)--(splitB)--node[below,note]{photon $B$}(eomB.west);
\draw[beam] (eomB.east)--(anaB.west);
\draw[beam] (anaB.east)--++(0.65,0)|-(Bp.west);
\draw[beam] (anaB.east)--++(0.65,0)|-(Bm.west);
\draw[signal] (Ap.east)--(tag.north west); \draw[signal] (Am.east)--(tag.north west);
\draw[signal] (Bp.east)--(tag.south west); \draw[signal] (Bm.east)--(tag.south west);
\draw[signal] (sensA.east)-|(tag.north);
\draw[signal] (sensB.east)-|(tag.south);
\draw[signal] (rng.north east)--(piezoA.south);
\draw[signal] (rng.south east)--(piezoB.north);
\draw[signal] (rng.east)--(tag.west);
\draw[cyan!60!black,very thick,-{Latex[length=2.5mm]}]
 (piezoA.west)--(movingA.east);
\draw[cyan!70!black,very thick,{Latex[length=2.3mm]}-{Latex[length=2.3mm]}]
 (piezoB.west)--(movingB.east);
\draw[orange!80!black,very thick,-{Latex[length=2.5mm]}]
 ($(movingA.north west)+(-0.1,0.1)$) arc[start angle=210,end angle=510,x radius=6mm,y radius=4mm];
\draw[orange!80!black,very thick,-{Latex[length=2.5mm]}]
 ($(movingB.south west)+(-0.1,-0.1)$) arc[start angle=150,end angle=450,x radius=6mm,y radius=4mm];
\node[note,orange!80!black] at (4.2,4.25) {$\phi_A\simeq\phi_B$};
\node[note,cyan!60!black] at (9.2,3.75) {$\vct a_A\simeq\vct a_B$};
\end{tikzpicture}}
\caption{Two-arm photon experiment with controllable motion correlation.  The
two analysis modules can undergo synchronous, partially correlated,
independent, or differential motion.  Mode II applies equivalent synchronous optical control
with EOM A and EOM B.  In Mode III both modules rotate with
$\phi_A(t)\simeq\phi_B(t)$; in Mode IV two feedback-controlled piezoelectric
stages apply the same seeded multiaxial vibration, so that
$\vct a_A(t)\simeq\vct a_B(t)$.  Independent encoders and accelerometers
measure the common variables $\phi_+$ and $\vct a_+$ and the residual
differential variables $\phi_-$ and $\vct a_-$.  A single rigid moving
platform carrying both modules gives the experimentally simpler common-frame
limit, whereas the two-stage arrangement permits the motion correlation
between the arms to be varied deliberately.
Solid red lines denote optical paths and dashed blue lines synchronization and
measurement signals.}
\label{fig:experiment}
\end{figure}

The independently driven two-stage arrangement is more general than a single
moving platform because it permits the correlation of the motion in the two
arms to be varied in a controlled way.  For each driven component one may
introduce the normalized cross-correlation
\begin{equation}
 \rho_k(\omega)=
 \frac{\operatorname{Re}S_{a_{A,k}a_{B,k}}(\omega)}
 {\sqrt{S_{a_{A,k}a_{A,k}}(\omega)
 S_{a_{B,k}a_{B,k}}(\omega)}},
 \qquad -1\leq\rho_k\leq1.
 \label{eq:motion_correlation_coefficient}
\end{equation}
Thus $\rho_k\simeq1$ realizes common motion, $\rho_k\simeq0$ realizes
independent motion, and $\rho_k\simeq-1$ realizes an oppositely driven
differential control.  A single rigid platform carrying both analysis modules
is the experimentally simpler common-frame limit, for which
$\vct a_-\simeq0$ and $\phi_-\simeq0$.  The two-stage configuration shown in
Fig.~\ref{fig:experiment} is retained because it allows continuous
interpolation between common, partially correlated, independent, and
differential motion.

The linear translational response is written
\begin{equation}
 \delta\Corr_{ij}(\omega)=
 \sum_{k=x,y,z}\left[
 \chi^{(a,+)}_{ijk}(\omega)a_{+,k}(\omega)
 +\chi^{(a,-)}_{ijk}(\omega)a_{-,k}(\omega)\right],
 \label{eq:multiaxial_acceleration_response}
\end{equation}
where $\chi^{(a,+)}$ is the sought susceptibility to the common non-inertial
motion and $\chi^{(a,-)}$ quantifies differential motion and local artifacts.
Different common drive directions probe different components of the
non-inertial susceptibility. With
$\vct z=(\vct a_+,\vct a_-)^{\mathsf T}$, both susceptibilities are estimated
from the correlation--acceleration cross-spectral matrix,
\begin{equation}
 \bigl(\vct\chi^{(a,+)}_{ij},\vct\chi^{(a,-)}_{ij}\bigr)
 =\vct S_{C_{ij}z}(\omega)\,\vct S_{zz}^{-1}(\omega).
 \label{eq:cross_spectral_susceptibility}
\end{equation}
The seeded waveform makes repeated runs and null tests directly comparable,
whereas the two accelerometers supply the physical common and differential
inputs used in the analysis.

\subsection{Phase-synchronous reconstruction and controls}

Each event is assigned the phase
\begin{equation}
 \varphi_\Omega=\Omega t\pmod{2\pi}.
\end{equation}
For $M$ phase bins, the first harmonic estimator is
\begin{equation}
 \widehat\Corr_{ij}^{(1)}=\frac2M\sum_{m=1}^M
 \Corr_{ij}(\varphi_m)e^{-i\varphi_m}.
\end{equation}
Four modes are compared: (I) stationary baseline; (II) stationary synchronous
optical modulation of both analyzers reproducing the same basis
transformations; (III) common harmonic rotation with
$\phi_A(t)\simeq\phi_B(t)$; and (IV) common seeded multiaxial piezoelectric
vibration with $\vct a_A(t)\simeq\vct a_B(t)$. A deliberately imposed
differential drive provides an additional calibration of local artifacts.
The optical--mechanical comparison is supplemented by two-sided kinematic
compensation,
\begin{equation}
 \widetilde\Corr(t)=R_A^{-1}(t)\Corr_{\rm meas}(t)R_B^{-\mathsf T}(t),\qquad
 \Delta\Corr_{\res}(t)=\widetilde\Corr(t)-C^{(0)}.
\end{equation}
Reversing the rotation direction tests
\begin{equation}
 \eta_\gamma(-\Omega)=\eta_\gamma(\Omega),\quad
 q_\gamma(-\Omega)=q_\gamma(\Omega),\quad
 \kappa_\gamma(-\Omega)=-\kappa_\gamma(\Omega).
\end{equation}
In Mode IV the principal observables are the cross spectra with both
$\vct a_+$ and $\vct a_-$, their full autospectral matrix, and the
corresponding magnitude-squared and partial coherences. A common-frame
candidate must follow $\vct a_+$ after conditioning on $\vct a_-$; a response
that follows only $\vct a_-$ is classified as differential motion or a local
artifact. Repeating the same seeded waveform with the
optical beam blocked, with separable photons, and with a mechanically isolated
dummy payload identifies electronic, detector, and mounting artifacts.

For $N_{ij}$ detected pairs, a leading statistical estimate is
\begin{equation}
 \sigma(C_{ij})\simeq\sqrt{\frac{1-C_{ij}^2}{N_{ij}}},
\end{equation}
while a full analysis should propagate the Poisson uncertainties of the four
coincidence counts, including accidental coincidences and dark counts. The
minimum resolvable susceptibility scales as
\begin{equation}
 \chi_{ij}^{\min}\sim
 \frac{\sigma(\widehat\Corr_{ij}^{(1)})}{\xi_0}.
\end{equation}

Principal systematic effects include analyzer-angle error, unintended angular motion,
optical-path modulation, detector-efficiency modulation, source drift,
accidental coincidences, decoherence, and thermal drift. Single counts,
two independent encoders, two three-axis accelerometers, beam-position monitoring,
shifted coincidence windows, interleaved
control runs, and repeated state tomography are therefore required. For an
ideal compensated $\ket{\Phi^+}$ state, the standard null prediction is
\begin{equation}
 \widetilde\Corr^{(\gamma)}(t)=\operatorname{diag}(1,1,-1),\qquad
 \eta_\gamma(\Omega)=q_\gamma(\Omega)=\kappa_\gamma(\Omega)=0.
\end{equation}

\section{Discussion}

The distinction
$E_\gamma=+\cos\theta$ versus $E_f=-\cos\theta$ concerns the sign and
geometry of the photon and fermion correlation sectors. It does not alter
Bell's theorem and does not assume that
non-inertial motion necessarily changes quantum nonlocality. Its purpose is to
distinguish a change of scalar projections from a change of the reconstructed
correlation object. A variation of $E$ or $S$ alone may result entirely from a
rotation of local bases. Singular values of the compensated correlation matrix
and an independently reconstructed density matrix provide stronger tests.

Three outcomes are possible: a purely kinematic response; a residual fully
accounted for by standard optical or relativistic transformations; or an
additional reproducible residual. Only the third requires an extended
dynamical interpretation. A candidate signal must be phase locked or
spectrally coherent with the measured motion, reproducible in independent
runs, scale consistently with $\phi_0$, $\Omega$, and the acceleration
spectrum, transform correctly under rotation reversal and changes of vibration
direction, be coherent with the common variables $\phi_+$ or $\vct a_+$ but
not explainable by the measured differential variables $\phi_-$ or
$\vct a_-$, and remain after all
calibrated optical, mechanical, and relativistic backgrounds are included.
Common motion of both analysis arms is therefore the primary test of a shared
non-inertial variable. Single-arm or deliberately mismatched motion is retained
as a differential control, not interpreted as the sought common-frame
response. Standard quantum theory predicts that an exactly common calibrated
rotation cancels from the relative analyzer angle and that the compensated
tensor remains unchanged; a nonzero common residual is consequently a
falsifiable hypothesis rather than an assumed effect.

\section{Conclusion}

The photon correlation obtained by averaging two projection amplitudes over
the complete phase interval $0\leq\lambda<2\pi$ is
$M_{AB}^{(\gamma)}(a,b)=+\cos(a-b)$. For physical linear-polarizer axes this
becomes $E_\gamma=+\cos2(\theta_A-\theta_B)$. The fermionic singlet remains
$E_f=-\cos(\alpha_A-\alpha_B)$ because the complete phase--momentum exchange
loop carries the negative holonomy. The full phase interval and the projective
periodicity of linear polarization are therefore distinct geometric elements
and must not be conflated.

The associated unbiased binary probabilities reproduce these signed cosine
laws and lead to opposite signed optimal CHSH combinations with the same
absolute Tsirelson value. The non-inertial construction therefore remains
inside the quantum correlation bound while retaining an operational distinction
between the two exchange sectors.

A Bell observable is a scalar projection of a complete two-particle
correlation tensor. For the photon state $\ket{\Phi^+}$ the baseline Stokes
block is $C^{(\gamma,0)}=\operatorname{diag}(1,1,-1)$, whereas for the
fermionic singlet $C^{(f,0)}=-I$. For a modulated system we introduced the
extended tensor $\FullT_{\mu\nu}(\vct\xi)$ and its susceptibility
$X_{\mu\nu\alpha}(\Omega)$. Phase-synchronous Stokes tomography, comparison
with equivalent optical modulation, common harmonic rotation and matched
multiaxial piezoelectric vibration of both arms, explicit common--differential
decomposition, and two-sided kinematic compensation define complementary null
tests for an additional non-inertial response. The construction replaces
isolated scalar variations by a reconstructible dynamical tensor and provides
an experimentally testable framework for Bell correlations in non-inertial
reference frames.


\begin{thebibliography}{99}
\bibitem{Bell1964} J.~S. Bell, On the Einstein Podolsky Rosen paradox,
\textit{Physics} \textbf{1}, 195--200 (1964).
\bibitem{CHSH1969} J.~F. Clauser, M.~A. Horne, A.~Shimony, and R.~A. Holt,
Proposed experiment to test local hidden-variable theories,
\textit{Phys. Rev. Lett.} \textbf{23}, 880--884 (1969).
\bibitem{Cirelson1980} B.~S. Cirel'son,
Quantum generalizations of Bell's inequality,
\textit{Lett. Math. Phys.} \textbf{4}, 93--100 (1980).
\bibitem{LaidlawDeWitt1971} M.~G.~G. Laidlaw and C.~M. DeWitt,
Feynman functional integrals for systems of indistinguishable particles,
\textit{Phys. Rev. D} \textbf{3}, 1375--1378 (1971).
\bibitem{Horodecki1995} R.~Horodecki, P.~Horodecki, and M.~Horodecki,
Violating Bell inequality by mixed spin-$\tfrac12$ states,
\textit{Phys. Lett. A} \textbf{200}, 340--344 (1995).
\bibitem{GingrichAdami2002} R.~M. Gingrich and C.~Adami,
Quantum entanglement of moving bodies,
\textit{Phys. Rev. Lett.} \textbf{89}, 270402 (2002).
\bibitem{PeresTerno2004} A.~Peres and D.~R. Terno,
Quantum information and relativity theory,
\textit{Rev. Mod. Phys.} \textbf{76}, 93--123 (2004).
\bibitem{TerashimaUeda2003} H.~Terashima and M.~Ueda,
Relativistic Einstein--Podolsky--Rosen correlation and Bell's inequality,
\textit{Int. J. Quantum Inf.} \textbf{1}, 93--114 (2003).
\end{thebibliography}
\end{document}